\documentclass[aps,prb,twocolumn,showpacs,superscriptaddress,groupedaddress]{revtex4}  
\usepackage{graphicx}  
\usepackage{dcolumn}   
\usepackage{bm}        
\usepackage{amssymb}   
\usepackage{amsmath}
\usepackage{epstopdf}
\usepackage{ragged2e}

\newcommand{\bx}{\bold{x}}
\newcommand{\bk}{\bold{k}}
\newcommand{\bbr}{\bold{r}}
\newcommand {\be}{\begin{eqnarray}}
\newcommand {\ee}{\end{eqnarray}}  
\newcommand {\ba}{\be \begin{aligned}} 
\newcommand {\ea}{\end{aligned} \ee}

\begin{document}



\title{Emergence of superconductivity in a doped single-valley quadratic band crossing system of spin-1/2 fermions}
\author{Kelly Ann Pawlak} \affiliation{University of California, Santa Barbara, California 93106, USA}\affiliation{National High Magnetic Field Laboratory and Department of Physics,
Florida State University, Tallahasse, Florida 32306, USA}
\author{James M. Murray}\affiliation{National High Magnetic Field Laboratory and Department of Physics,
Florida State University, Tallahasse, Florida 32306, USA}
\author{Oskar Vafek} \affiliation{National High Magnetic Field Laboratory and Department of Physics,
Florida State University, Tallahasse, Florida 32306, USA}

\date{\today}

\begin{abstract}
\noindent For two-dimensional single-valley quadratic band crossing systems with weak repulsive electron-electron interactions, we show that upon introducing a chemical potential, particle-hole order is suppressed and superconductivity becomes the leading instability. In contrast to the two-valley case realized in bilayer graphene, the single-valley quadratic band touching is protected by crystal symmetries, and the different symmetries and number of fermion flavors can lead to distinct phase instabilities. Our results are obtained using a weak-coupling Wilsonian renormalization group procedure on a low-energy effective Hamiltonian relevant for describing electrons on checkerboard or kagom\'e lattices. In 4-fold symmetric systems we find that $d$-wave and $s$-wave superconductivity are realized for short-ranged (Hubbard) and longer-ranged (forward scattering), respectively. In the 6-fold symmetric case, we find either $s$-wave superconductivity or no superconducting instability. 
\end{abstract}

\maketitle


\section{Introduction}
Recently, two-dimensional quadratic band crossing (QBC) systems, in which two parabolically dispersing bands meet at a single point in momentum space, have emerged as an attractive venue in which to study multicriticality and competing orders.\cite{KaiSun,vafekyang,Nandkishore,Zhang,Throck,ChernBatista,Lemonik,Vafek,Murray,Cook,KaisunCold} In contrast to systems with Dirac points in two spatial dimensions, which are robust to weak short-range electron-electron interactions, analogously constructed QBC systems are prone to instabilities even at weak coupling. Indeed, arbitrarily weak interactions can lead to non-trivial electronic phases in QBC systems.  In a recent paper by two of us \cite{Murray} it was found that a single-valley QBC system, with a single quadratic band touching point protected by point group and time reversal symmetries, displays spontaneous symmetry breaking towards topological insulating phases when the chemical potential lies precisely at the band degeneracy point. These results hold for bandstructures arising on the checkerboard and kagom\'e lattices, which have $C_{4v}$ and $C_{6v}$ symmetry, respectively. Unlike in the AB stacked honeycomb bilayer with $D_{3d}$ symmetry,\cite{TrigWarpingFootnote} such QBCs are symmetry-protected and lack valley degeneracy, making them an important model case to consider due to their robustness and relative simplicity.

In the case of the AB stacked honeycomb bilayer, moving away from half filling, it has recently been shown that repulsive electron-electron interactions generically lead to superconducting orders.\cite{Vafek,MurrayLong} The appearance of such an unconventional superconducting phase adjacent to a particle-hole ordered phase is a ubiquitous phenomenon in several families of strongly correlated materials, and QBC systems provide a uniquely tractable window through which such competing (or ``intertwined'') orders can be understood.  As pointed out previously\cite{vafekyang,Lemonik}, the number of fermion flavors can affect the nature of the instability, so there is no reason to expect the results already found for the two-valley case of honeycomb bilayer should carry over to the single-valley case. 
In this paper we extend previous results \cite{Murray} by investigating the single-valley QBC system with $C_{4v}$ or $C_{6v}$ symmetry away from half filling, showing that $d$-wave and conventional $s$-wave superconducting order can be realized at nonzero doping for repulsive short-range and longer-range interactions, respectively.

 \section{Model and renormalization group equations}
 \label{sec:model}
\begin{figure}
\includegraphics[width=0.45\textwidth]{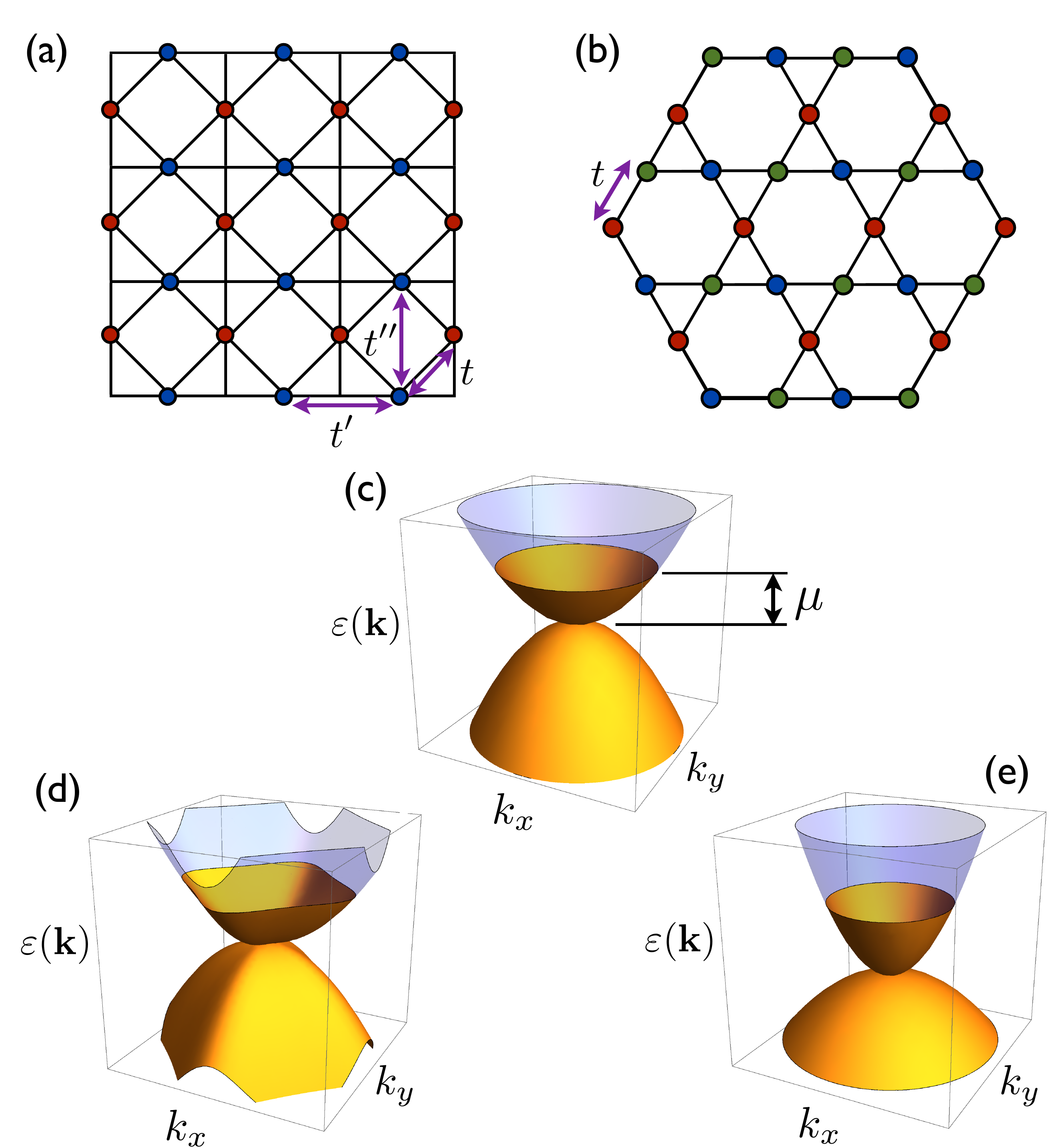}
\caption{(a) Checkerboard lattice, with nearest neighbor ($t$) and next-nearest neighbor ($t', t''$) hoppings shown. (b) Kagome lattice, with nearest neighbor hopping $t$. (c) Energy bands near a 2D quadratic band touching point in the rotationally invariant and particle-hole symmetric limit, with chemical potential $\mu$. (d) For $\eta = 1.3\frac{\pi}{4} \neq \frac{\pi}{4}$, the dispersion has only 4-fold rotational symmetry. (e) For $\lambda = 0.4 \neq 0$, the bands become particle-hole asymmetric.}
\label{fig:1111a}
\end{figure}

We use the following low-energy effective Hamiltonian to describe a two-dimensional QBC system : \cite{KaiSun}
\begin{align}
\label{eq:hamiltonian}
H=H_0+H_{int},
\end{align}
where
\ba
\label{eq:0108a}
H_0 &= \sum_{|\bk| < \Lambda} \sum_{\alpha=\uparrow\downarrow} \psi^\dagger_{\bk\alpha} \mathcal{H}_0 (\bk) \psi_{\bk\alpha}, \\
\mathcal{H}_0(\bk) &= t_I \bk^2 1 + 2t_x k_x k_y \sigma_1 + t_z (k_x^2 - k_y^2)\sigma_3 \\
&= \frac{\bk^2}{\sqrt{2} m} \left[ \lambda 1_2 + \sin \eta \sin 2\theta_\bk \sigma_1
 + \cos \eta \cos 2\theta_\bk \sigma_3 \right],
\ea
where $\Lambda$ is an ultraviolet momentum cutoff, ${\sigma_i}$ are the usual Pauli matrices with $i=0$ to be understood as the $2\times 2$ identity, $1/\sqrt{2}m = \sqrt{t_x^2 + t_z^2}$, $\lambda = t_I / \sqrt{t_x^2 + t_z^2}$ describes particle-hole anisotropy, and $\tan \eta = t_x / t_z$ describes rotational anisotropy. The corresponding dispersion is
\be
\label{eq:1012b}
\varepsilon_\pm (\bk) 
= \frac{\bk^2}{2m} \left[ \sqrt{2}\lambda \pm \sqrt{1+\cos(2\eta) \cos(4\theta_\bk)} \right].
\ee
For $|t_I| < \mathrm{min}(|t_x|,|t_z|)$, the model has two parabolic bands meeting at $\bk=0$, with one band dispersing upward and the other downward as one moves away from this point, as shown in Figure \ref{fig:1111a}.
In the rotationally invariant and particle-hole symmetric cases we have $\eta = \pi/4$ and $\lambda = 0$, respectively. The interaction term appearing in \eqref{eq:hamiltonian} is
\begin{align}
\label{eq:H_int}
H_\mathrm{int} = \frac{2\pi}{m} \sum_{i=0}^3 g_i \int d^2 x
	\left( \sum_{\alpha=\uparrow\downarrow}
	\psi^\dagger_\alpha (\bx) \sigma_i \psi_\alpha (\bx) \right)^2,
\end{align}
which contains all marginal symmetry-allowed couplings.

As written, the Hamiltonian \eqref{eq:hamiltonian}--\eqref{eq:H_int} is invariant under the symmetries of the $C_{4v}$ point group, and may arise as a low-energy effective theory for electrons on the checkerboard lattice at half filling, with the parameters in \eqref{eq:0108a} related to the lattice hopping parameters shown in Figure \ref{fig:1111a} as $t_x = t/2$, $t_z = (t' - t'')/2$, and $t_I = (t' + t'')/2$.\cite{KaiSun} It can also describe a system having $C_{6v}$ symmetry if one takes $t_x = t_z = t$ and $g_1 = g_3$. In this case the low-energy theory has full rotational symmetry, since any other possible terms consistent with six-fold rotational symmetry would contain higher powers of momentum and hence would be irrelevant. Such an effective theory may arise from electrons on a kagom\'e lattice at $\frac{2}{3}$ filling.\cite{KaiSun} In either case, the QBC point is robust to perturbations which do not break time reversal or point group symmetries. In the absence of external symmetry-breaking fields, only a spontaneous symmetry-breaking instability to an ordered phase can alter the QBC point. The group representations for the $C_{4v}$ and $C_{6v}$ cases together with the corresponding symmetry-allowed interaction terms are shown in Table \ref{table:matrices}. 
\begin{table}
\begin{center}
\begin{tabular}{lcccc}
\hline
\hline
Interaction & $g_0$ & $g_1$ & $g_2$ & $g_3$ 
\\ \hline
Forward scattering & $g$ & $0$ & $0$ & $0$ \\
\quad  (checkerboard or kagome) & & & & \\
Hubbard (checkerboard) & $\frac{1}{2}U$ & $0$ & $0$ & $\frac{1}{2}U$ \\
Hubbard (kagome) & $\frac{2}{3} U$ & $\frac{1}{3} U$ & $0$ & $\frac{1}{3} U$ \\
\hline
\hline
\end{tabular}
\end{center}
    \caption{Bare interactions $g_i(0)$ appearing in \eqref{eq:H_int} for longer-range (forward scattering) and short-range (Hubbard) interactions, as determined by deriving the low-energy effective theory from a microscopic model on the checkerboard or kagome lattice.}
    \label{table:interactions}
\end{table}
The contact interactions $g_i$ appearing in \eqref{eq:H_int} can be related to interactions in the lattice model by deriving the low-energy effective theory from a tight-binding model. As shown in Table \ref{table:interactions}, this mapping will be different in the $C_{4v}$ and $C_{6v}$ cases.
\begin{table*}
\begin{ruledtabular}
    \begin{tabular}{lllllllll}
 Rep. ($C_{4v}$) & $g_i$ & $M_i^{(c)}$ & Phase (c) & $M_i^{(s)}$ & Phase (s) & $ M^{(pp)}_i$ & $\tilde{g}_i$& Phase (pp)  \\ \hline
 $A_1$ & $g_0$ & $1_4$ & -- & $1 \vec{s}$ & FM & $1 s_2$ & $\tilde{g}_0$& $s$\\
 $A_2$ & $g_2$ & $\sigma_2 1$ & QAH & $\sigma_2 \vec{s}$ & QSH & $\sigma_2 (is_2\vec{s})$ &$\tilde{g}_2$& $p$ \\
 $B_1$ & $g_3$ & $\sigma_3 1$ & Nem.~(site) & $\sigma_3 \vec{s}$ & NSN (site) & $\sigma_3 s_2$ & $\tilde{g}_3$& $d_{x^2 - y^2}$ \\
 $B_2$ & $g_1$ & $\sigma_1 1$ & Nem.~(bond) & $\sigma_1 \vec{s}$ & NSN (bond) & $\sigma_1 s_2$ & $\tilde{g}_1$& $d_{xy}$\\
\hline
\hline
Rep. ($C_{6v}$) 
\\ \hline
 $A_1$ & $g_0$ & $1_4$ & -- & $1 \vec{s}$ & FM & $1 s_2$ & $\tilde{g}_0$& $s$\\
 $A_2$ & $g_2$ & $\sigma_2 1$ & QAH & $\sigma_2 \vec{s}$ & QSH & $\sigma_2 (is_2\vec{s})$ &$\tilde{g}_2$& $p$ \\
 $E_2$ & $(g_1,g_3)$ & $(\sigma_3 1, \sigma_1 1)$ & Nem. & $(\sigma_3 \vec{s}, \sigma_1 \vec{s})$ & NSN & $(\sigma_3 s_2, \sigma_1 s_2)$ & $(\tilde{g}_3, \tilde{g}_1)$& $(d_{x^2 - y^2}, d_{xy})$ \\
    \end{tabular}
    \end{ruledtabular}
    \caption{Fermionic couplings $g_i$, together with the representation of $C_{4v}$ or $C_{6v}$ under which they transform, the matrices appearing in the source term bilinears \eqref{eq:0206a}, and the phases associated with each bilinear. The possible excitonic phases are ferromagnet (FM), quantum anomalous Hall (QAH), quantum spin Hall (QSH), charge nematic on sites or bonds, and nematic-spin-nematic (NSN) on sites or bonds. The last three columns show the matrices appearing in the particle-particle bilinears in \eqref{eq:0206a}, the transformed couplings from \eqref{eq:1103a}, and the corresponding superconducting phases (s-wave, p-wave, and d-wave). In the left most columns the pair couplings $\tilde{g}_i$ are given with their corresponding superconducting phases.}
    \label{table:matrices}
\end{table*}

We employ a Wilsonian renormalization group (RG) procedure in order to study the effects of interactions and instabilities to ordered phases at low energy scales.\cite{Shankar} It is useful to define the following action:
\begin{align}
\label{eq:action}
S = \int d\tau \left[ \sum_{|\bk| < \Lambda} \sum_\alpha
	\psi^\dagger_{\bk\alpha}(\partial_\tau + \mathcal{H}_0 (\bk) )\psi_{\bk\alpha}
	+ H_\mathrm{int} \right],
\end{align}
where the Grassmann fields $\psi_{\bk\sigma}$ now depend on imaginary time $\tau$.
The RG step is then performed by eliminating states within the momentum shell $\Lambda(1-d\ell) < |\bk| < \Lambda$ while integrating over all frequencies. By including all one-loop diagrams and rescaling the couplings after each RG step, one obtains the following flow equations:
\begin{align}
\label{eq:0128a}
\frac{d g_i}{d\ell} = \sum_{j,k=0}^3 A_{ijk}g_j  g_k,
\end{align}
where the coefficients $A_{ijk} (\mu)$ are given by a sum over the five diagram contributions: \begin{align}
A_{ijk}(\mu) = \sum_{d=1}^5 A_{ijk}^d(\mu),
\end{align}
with details provided in the Appendix.
The parameters $\lambda$ and $\eta$ do not flow at this order; however, the chemical potential does indeed flow such that $\mu$ rescales as $\mu \rightarrow \mu e^{2\ell}$ at T=0.\cite{Vafek,MurrayLong} One sees that the couplings are marginally relevant and generally flow to infinite values for sufficiently large $\ell$, though their ratios approach fixed finite values, with these ratios ultimately determining the nature of the ordered phase. Due to the perturbative nature of our approach, the flow equations remain valid only at weak coupling and break down at RG scales where $|g_i(\ell)| \gtrsim 1$.  In much of what follows, we shall focus on the cases of short-range repulsive Hubbard interaction and longer-range forward scattering. The particular values of $g_i(0)$ for these two cases are shown in Table \ref{table:interactions}. These two interactions represent two extreme cases; one may tune to any intermediate range through a straightforward interpolation of the initial coupling values.

Away from half-filling, only particle-particle scattering leads to chemical potential dependence in the flow equations: 
\begin{align}
A_{ijk}^5(\mu_\ell) = \left( \frac{1}{1-\mu_\ell^2}\right) A_{ijk}^5(0) ,
\end{align}
where we have defined $\mu_\ell \equiv \mu e^{2\ell}/\frac{\Lambda^2}{2m}=\hat\mu e^{2\ell}$. Thus we may rewrite the flow equations such that the $\mu$-dependent terms are separate, in order to make use of the results established in Ref.~\onlinecite{Murray}:
\ba
\label{aijkform}
\frac{d g_i}{d\ell} &= \sum_{j,k=0}^3 \left(A_{ijk}(0) + \left( \frac{\mu_\ell^2}{1-\mu_\ell^2}\right) A_{ijk}^5(0)  \right)g_j  g_k
\\&= \dot{g_i}(\ell,\mu=0) + \left( \frac{\mu_\ell^2}{1-\mu_\ell^2}\right) \sum_{j,k=0}^3 A_{ijk}^5(0) g_j  g_k.
\ea
Further insight can be gained by using the so-called Fierz identities \cite{Vafek} to recast the interaction term as a combination of pairing interactions of the general form
\begin{align}
\label{eq:1103a}
S_{int}=\int \mathrm{d}\tau \int \mathrm{d}^2 r \left[ \sum_{j=\mathrm{singlet}} \tilde{g}_jS^\dagger_jS_j +\sum_{j=\mathrm{triplet}}\tilde{g}_j\vec{T}^\dagger_j \cdot \vec{T}_j \right],
\end{align}
where
\ba
S_j &=\psi^T(\bbr,\tau)\sigma_j s_2 \psi(\bbr,\tau), \; \; \; \text{for} \; j=0,1,3
\\\vec{T}_j &=\psi^T(\bbr,\tau)\sigma_2 \vec{s} \psi(\bbr,\tau), \; \; \; \text{for} \; j=2.
\ea
Here $\vec{s}$ is a vector of Pauli matrices corresponding to electron spin. Recasting the problem in this equivalent way and making use of the Fierz identities gives us the following relation among the ordinary and Cooper gauge couplings:
\begin{align}
\label{eq:Fierz}
\tilde{g_i} (\ell, \mu)= \sum_j \mathcal{F}_{ij} g_j (\ell, \mu)
\end{align}
where the Fierz matrix $\mathcal{F}$ is 
\begin{align} \mathcal{F} \equiv \frac{1}{4} \begin{pmatrix}
1 & 1 & -1 &1 \\
1 & 1 & 1 & -1 \\
1 & -1 & -1 & -1 \\
1 & -1 & 1 & 1
\end{pmatrix}
\end{align}
The importance of this transformation is that the flow equations now take on the form
\begin{align}
\label{eq18}
\frac{d\tilde{g}_i}{d\ell} =- \frac{\alpha_i}{1-\mu_\ell^2}\tilde{g_i}^2 + \sum_{j,k} \tilde{A}_{ijk} \tilde{g}_j ,\tilde{g}_k
\end{align}
where $\alpha_i \geq 0$, and the coefficients $\tilde{A}_{ijk}$ are non-singular as $\mu_\ell \to 1$. 
It is apparent that for $\mu_\ell \approx 1$ the equations will become essentially decoupled. In this regime, if a pair coupling is attractive, it will diverge to negative infinity while the repulsive couplings saturate. Following the arguments given in Ref.~\onlinecite{Vafek}, one can show that for arbitrarily weak couplings an appropriate $\hat\mu$ can always be chosen to accomplish this. First consider the case at half filling: at some $\ell_1$ one of the couplings becomes attractive. This behavior is approximately preserved as one moves away from half filling so long as $\mu_{\ell_1} \ll 1 $. On the other hand, we need to choose $\hat\mu$ sufficiently large such that the attractive coupling diverges at $\ell_{FS}$ defined by $\mu_{\ell_{FS}} = 1$ while saturating the repulsive couplings. If at half-filling, the repulsive couplings diverge at $\ell_*$, then having $\mu_{\ell_{FS}} \ll \mu_{\ell_*}$ would satisfy this. Combining these conditions produces the following inequality:
\be
\label{eq:1106a}
\frac{\Lambda^2}{2m} e^{-2\ell_*} \ll \mu \ll \frac{\Lambda^2}{2m} e^{-2\ell_1} .
\ee
In order to show that this relationship may always be satisfied, we defer to the following argument. Consider that the $\mu=0$ flow equations are invariant under the following transformation:
\begin{align}
g_i \rightarrow b g_i, \hspace{3mm} \ell  \rightarrow \ell/b.
\end{align}
Then we can say that there are constants $C_1$ and $C_*$ such that $\ell_1 = C_1/g$ and $\ell_*=C_*/g$, for $g \equiv \sqrt{\sum_i g_i^2(0) } $. Therefore, we have that
\begin{align}
\label{eq:1104a}
\frac{\Lambda^2}{2m} e^{-2\frac{C_*}{g}} \ll \mu \ll \frac{\Lambda^2}{2m} e^{-2\frac{C_1}{g}}
\end{align}
It is clear that as long as the couplings can be arbitrarily weak, the relationship can be satisfied. We use the above relation to determine the coupling strength and Fermi level appropriate to obtain a superconducting phase. For example, in the case of forward scattering at half filling, we find that $C_1\approx 0.26$ and $C_* \approx 0.40 $ (see Figure \ref{fig:graphs}). Choosing the coupling to be $g_0(0) = 0.05$, we find that  $\hat\mu = 10^{-6}$ entirely satisfies our constraint \eqref{eq:1104a}.

In order to determine the phase instabilities as the couplings grow large, we calculate the susceptibilities of the couplings at finite $\mu$ by introducing symmetry breaking terms coupled to source fields $\Delta_i^{(c,s,pp)}$ into the action:
\begin{align}
\label{eq:0206a}
\nonumber & S_\Delta =\int d\tau \int d^2 x \bigg\{ \sum_{i=1}^4 \bigg[ \Delta_i^{(c)} \psi^\dagger M_i^{(c)} \psi
	+ \vec{\Delta}_i^{(s)} \cdot \psi^\dagger \mathbf{M}_i^{(s)} \psi \bigg] \\
	&+ \frac{1}{2} \bigg[ \sum_{i=1}^3 \Delta^{(pp)}_i \psi^\dagger M^{(pp)}_i \psi^* 
	+ \vec{\Delta}^{(pp)}_4 \cdot \psi^\dagger \mathbf{M}^{(pp)}_4 \psi^* + H.c. \bigg] \bigg\}.
\end{align}
The matrices that define the various fermion bilinears in charge (c), spin (s), and particle-particle (pp) channels are given in Table \ref{table:matrices}.

In order to further investigate the $\mu$-dependence of the instabilities, we can look at the power-law behavior of these susceptibilities near $\ell^*$, which are of the form $\chi_i \sim (\ell^* - \ell)^{-\gamma_m}$, where $\gamma_m$ can be shown to be\cite{gammaFootnote}
\begin{align}
\label{eq:1103b}
\gamma^{(c,s,pp)}_m (\mu) = \frac{2 \sum_j B_{mj}^{(c,s,pp)}\rho_j}{\sum_{ijk}A_{ijk} \rho_j \rho_k \rho_i} -1 
\end{align}
for $\rho_i \equiv \lim_{\ell \rightarrow \ell^*} g_i(\ell)/\sqrt{\sum_j g_j^2(\ell)}$. The $B_{ij}$'s as well as the equations for the susceptibilities themselves are defined in the Appendix.
These values determine the rate at which each susceptibility grows exponentially near the Fermi surface, therefore serving as an indicator of the leading order. As one can see from the plots given in Figure \ref{fig:gamma}, SC order is not realized until the system is pushed past a critical value of $\hat\mu$. There is then a regime where SC order of the kind described in this paper remains the leading order. 

\section{Flow equation solutions}

\subsection{Isotropic case}
\begin{figure}
\includegraphics[width=0.48\textwidth]{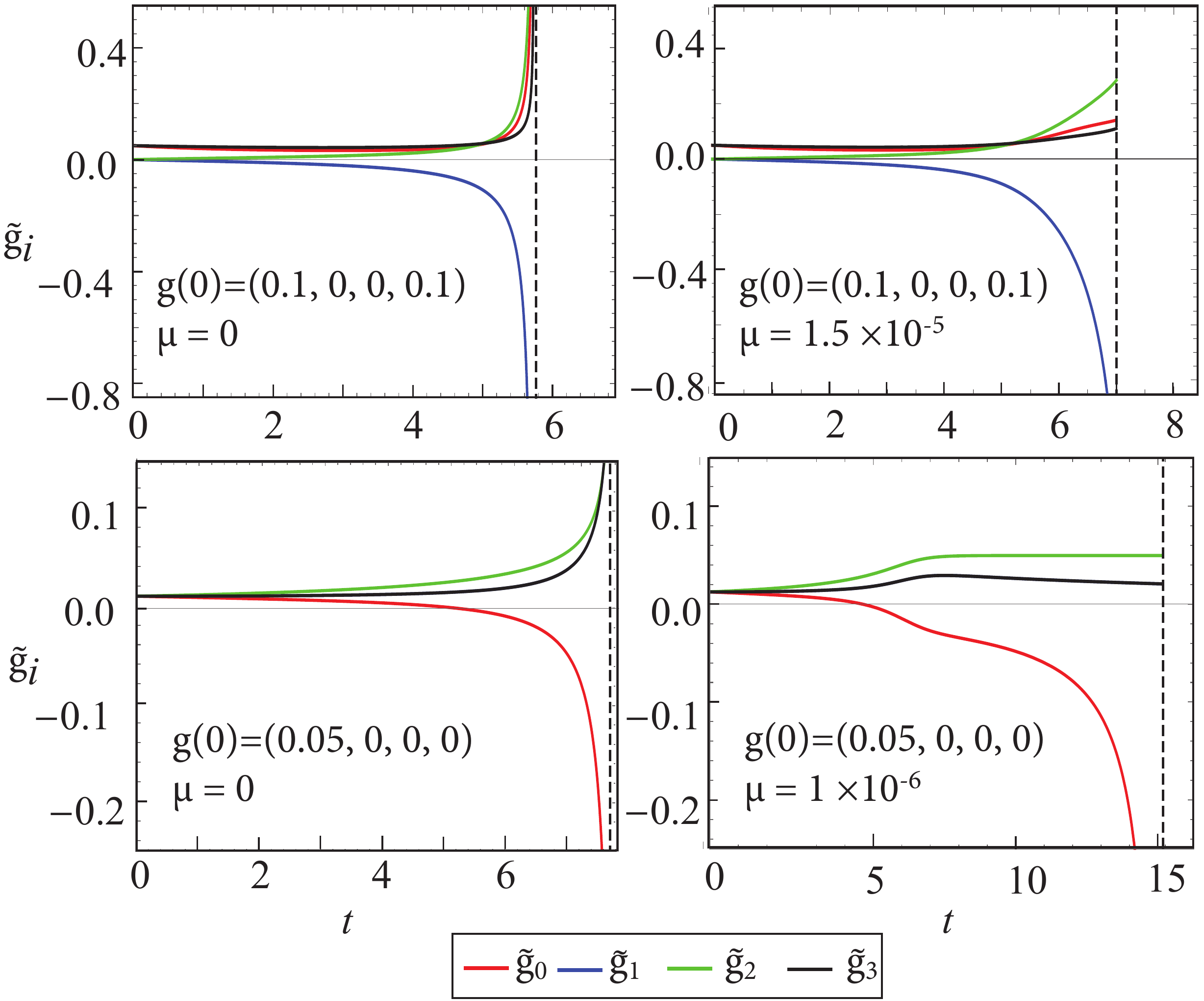}
\caption{Top row: Flows of couplings for Hubbard interaction (on the checkerboard lattice), with $\hat \mu = 0$ (left), and $\hat\mu = 1.5\times 10^{-5}$ (right). Bottom row: RG flows of couplings at half filling for forward scattering interaction with $\hat\mu = 0$ (left) and at $\hat\mu = 10^{-6}$ (right). In both cases, a sufficiently large chemical potential for a given interaction strength causes all repulsive couplings to saturate, while attractive couplings diverge. }\label{fierz}
\label{fig:graphs}
\end{figure}

We begin by analyzing the flow equations in the simplest case of particle-hole symmetry ($\lambda=0$) and rotational invariance ($\eta = \frac{\pi}{4}$). The flow equations \eqref{aijkform} in this case at half-filling ($\mu = 0$) are
\begin{equation}
\begin{aligned}
\label{eq:g_flows}
\dot{g}_0&= -4 g_0 g_+ , \\
\dot{g}_+&= -(g_0 - g_+)^2 - (g_2 - g_+)^2  - 6g_+^2 , \\
\dot{g}_2&= 4(g_0 g_2 - g_2^2 - g_-^2 + g_+^2 - 3 g_2 g_+) , \\
\dot{g}_-&= 2 g_-(g_0 - 3g_2 - 2g_+),
\end{aligned}
\end{equation}
where we have defined $g_\pm = \frac{1}{2} \left( g_3 \pm g_1 \right)$ to simplify the form of the equations. In the case of $C_{6v}$ symmetry, we have $g_-=0$, reducing the system to one with only three couplings. Results pertaining to these flow equations at half filling are given in Ref.~\onlinecite{Murray}. The $\mu$-dependent contributions to the flow equations come from the last term in \eqref{aijkform} and are given by
\ba
&\sum_{j,k=0}^3 A_{0jk}^5(0) g_j  g_k \\
	& \quad\quad\quad = g_0^2 + g_2^2 +2\left(g_+^2 +g_-^2+ g_+(g_0 - g_2)  \right) , \\
&\sum_{j,k=0}^3 \frac{1}{2}[A_{3jk}^5(0) + A_{1jk}^5(0) ] g_j  g_k =(g_0 -g_2 +2g_+)^2 , \\
&\sum_{j,k=0}^3 A_{2jk}^5(0) g_j  g_k  = 2\left(g_-^2 + (g_2-g_+)(g_0+g_+)\right) , \\
&\sum_{j,k=0}^3 \frac{1}{2}[A_{3jk}^5(0) - A_{1jk}^5(0) ] g_j  g_k = 4g_-\left(g_0 + g_2\right) .
\ea
It is straightforward to obtain the flows of the transformed couplings $\tilde g_i (\ell)$ using \eqref{eq:Fierz} from these equations. In Figure \ref{fig:graphs} we show the flow of the superconducting Fierzed couplings $\tilde g_i$ as plotted parametrically against a variable $t$ defined as \cite{Vafek}
\begin{align}
t= \frac{1}{2} \ln\left(\frac{1 - \hat\mu}{ e^{-2\ell} -\hat\mu}\right),
\end{align}
which vanishes at $\ell = 0$ and increases without bound as the Fermi surface is approached, thus better showing the behavior of the quickly diverging coupling flows as $\ell\to\ell_{FS}$. At half filling, the couplings diverge to both positive and negative
values, as shown in Figure \ref{fig:graphs}. 
Including a nonzero chemical potential gives an advantage to the attractive couplings, as is apparent from \eqref{eq18}, so that these diverge while the repulsive couplings saturate as the UV cutoff approaches the Fermi level. For forward scattering, in which case all initial couplings $\tilde{g}_i(0)$ are positive, running RG with optimal chemical potential and coupling strength causes the Fierz coupling associated with conventional s-wave SC to diverge to large negative values. In the case of the short-range Hubbard interaction two Fierz couplings start out at zero -- one of which, corresponding to a $d_{xy}$ SC state, diverges to large negative values under RG. 

\begin{figure}
\includegraphics[width=0.48\textwidth]{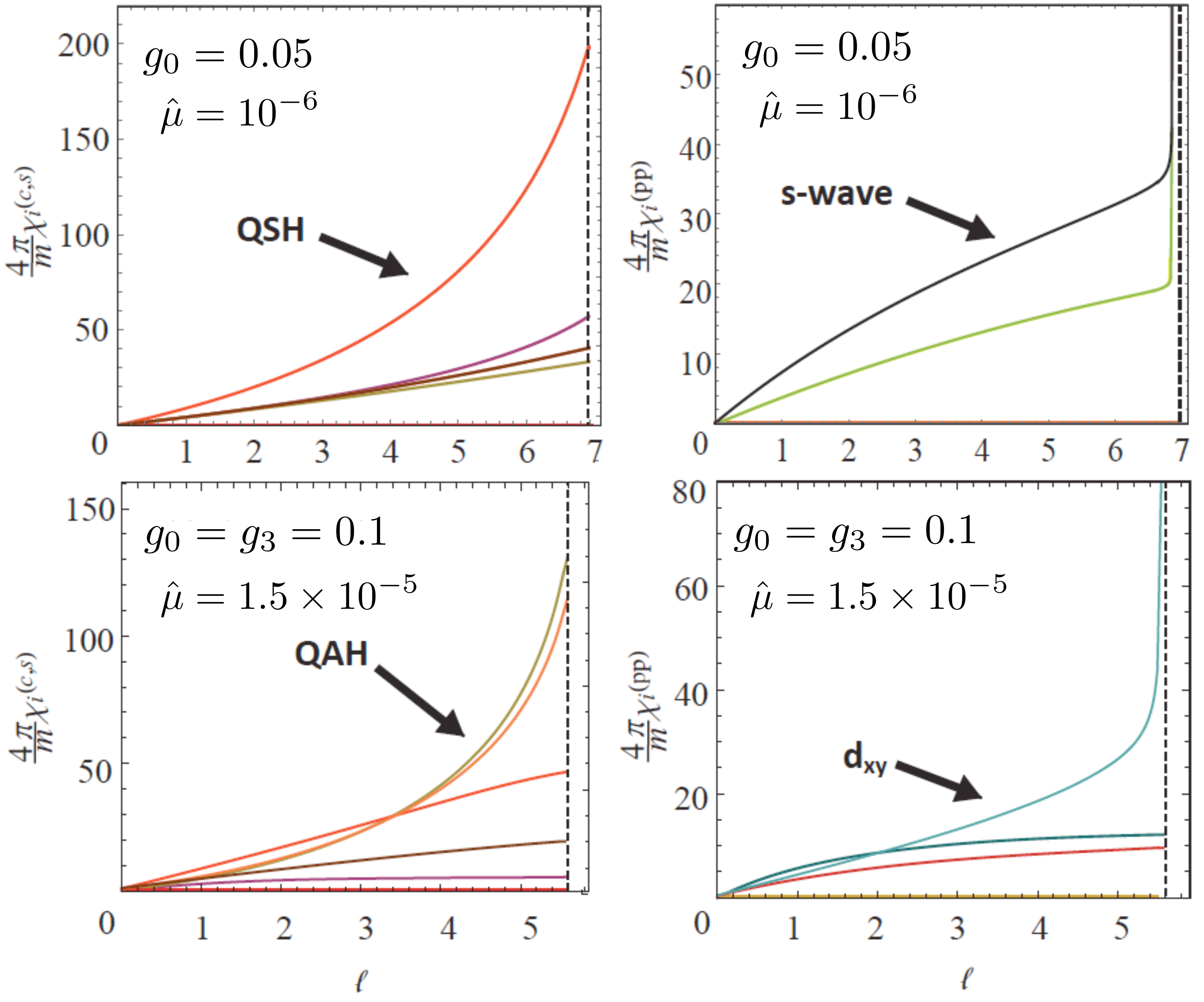}
\caption{RG flows of susceptibilities for various instabilities in particle-hole and particle-particle channels with finite chemical potential. Top row: With forward scattering interaction, susceptibilities in spin and charge channels saturate at finite values (left), while the susceptibility in the $s$-wave particle-particle channel diverges (right). Bottom row: With Hubbard interaction (on the checkerboard lattice), susceptibilities in spin and charge channels saturate at finite values (left), while the susceptibility in the $d$-wave particle-particle channel diverges (right).}
\label{foo}
\label{fig:suc}
\end{figure}

As shown in Figure \ref{fig:suc}, the instabilities at and near half-filling\cite{Murray} are to quantum anomalous Hall\cite{Haldane,Chang} (QAH) and quantum spin Hall\cite{Kane,Bernevig,Konig} (QSH) phases, both of which are topological in nature and feature charge or spin edge currents. (In the QSH case, edge spin currents will not in general be conserved in the presence of disorder.)
For larger $\mu$, the only susceptibilities that show divergent behavior for sufficiently large $\mu$ are those in the particle-particle channels, and in this case $s$-wave and $d$-wave superconducting phases are the leading instabilities of the doped system. The appearance of $s$-wave superconductivity driven by repulsive interaction is unusual, and comes about in this case due to the fact that (i) all couplings $\tilde g_i(0)$ are initially equal for longer-range interactions, so that no channel is initially disfavored; and (ii) the $s$-wave phase is fully gapped, making it favorable due to the increased gain in condensation energy. This is analogous to the pair density wave (PDW) instability previously found for repulsively interacting electrons on the honeycomb lattice.\cite{Vafek,MurrayLong} The PDW is also fully gapped, though in that case the Cooper pairs have nonzero total momentum $2\mathbf{K}$ due to the fact that pairing occurs within a Fermi pocket centered at wavevector $\mathbf{K}$.
\begin{figure}
\includegraphics[width=0.45\textwidth]{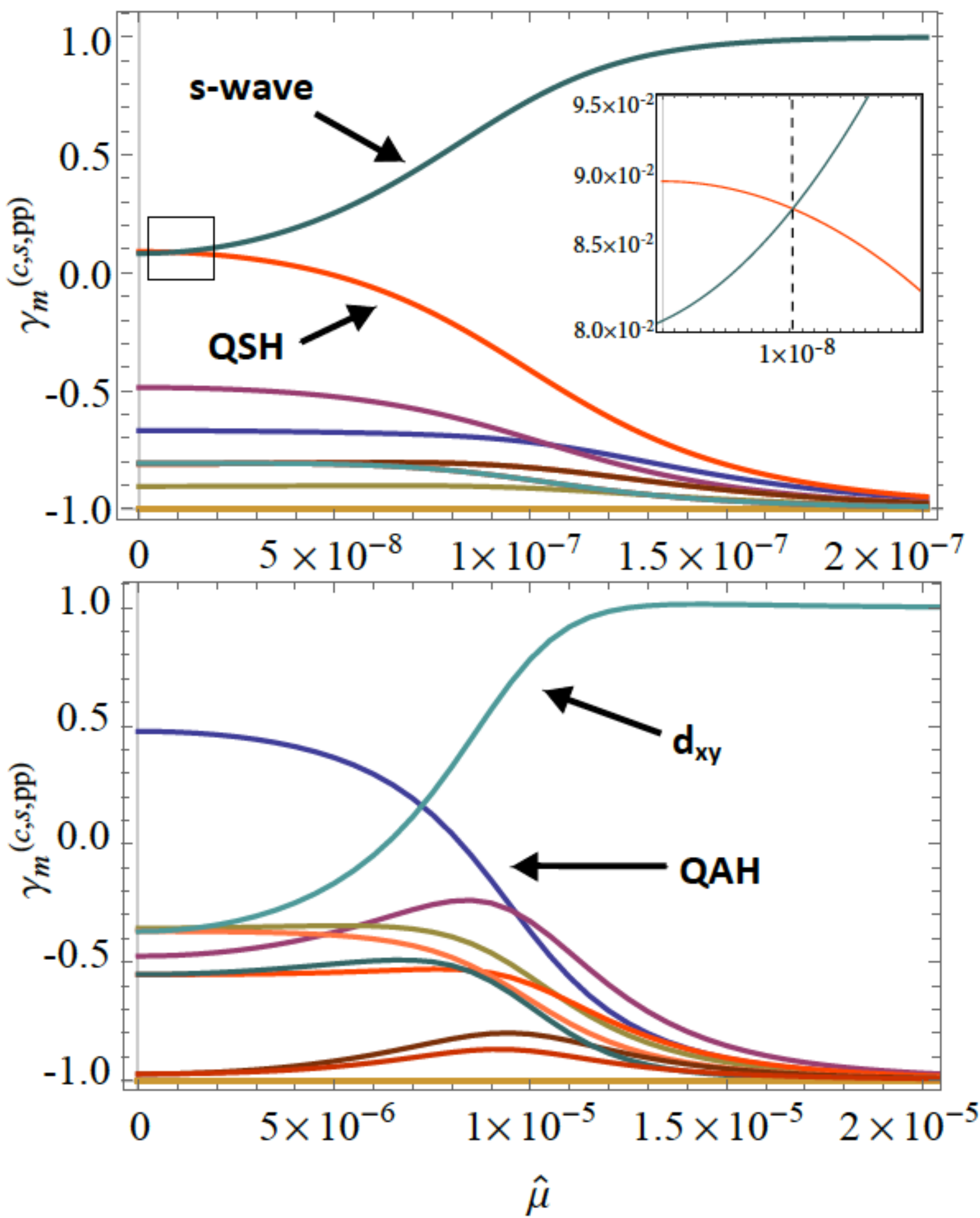}
\raggedright 
\caption{Variation of the susceptibility exponent $\gamma_m^{(c,s,pp)}$ with increasing dimensionless chemical potential $\hat\mu$. Moving away from half-filling causes the susceptibilities of the leading particle-hole instabilities to saturate, while those of the particle-hole channels diverge. Top: forward scattering interaction, with $g_0(0) = 0.05$ and $g_{1,2,3}(0) = 0$. Bottom: Hubbard interaction (on checkerboard lattice), with $g_0(0)=g_3(0)=0.1$ and $g_1(0)=g_2(0)=0$. }\label{foo}
\label{fig:gamma}
\end{figure}

\subsection{Anisotropic case}
\begin{figure}
\includegraphics[width=0.45\textwidth]{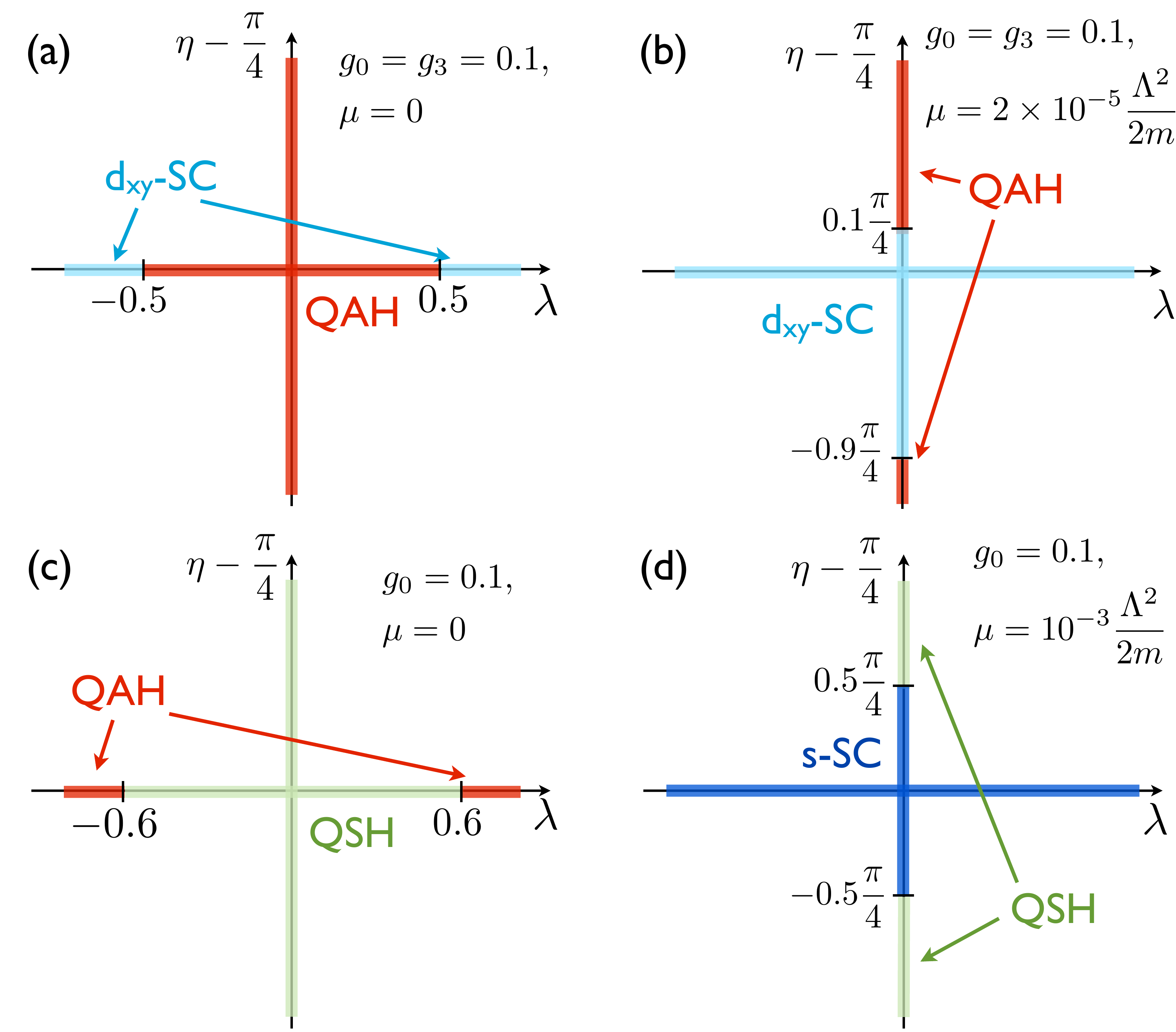}
\raggedright 
\caption{Phase instabilities at various values of angular anisotropy $\eta-\frac{\pi}{4}$ and particle-hole asymmetry $\lambda$. Top row: Hubbard interaction (for a checkerboard lattice) at zero doping (a) and with doping (b), with instabilities to quantum anomalous Hall (QAH) and $d$-wave superconducting ($d$-SC) phases. Bottom row: longer-ranged forward scattering interaction at zero doping (c) and with doping (d), with instabilities to QAH, quantum spin Hall (QSH), and $s$-wave superconducting ($s$-SC) phases.}
\label{fig:phasediagram}
\end{figure}

Let us now generalize the discussion by moving away from the rotationally invariant and particle-hole symmetric limit by including arbitrary $\eta$ and $\lambda$ in the flow calculations, which is done explicitly in the second portion of the Appendix. Solving these modified RG equations leads to the phase diagrams shown in Figure \ref{fig:phasediagram}. In calculating these phase diagrams, the leading instabilities are determined by taking the largest susceptibility when the couplings attain values $|g_i(\ell)| \gtrsim 1$. Another possible criterion for determining phases is to take the largest susceptibility exponent $\gamma$ where the flows diverge. The phase diagrams resulting from this choice are qualitatively similar to those shown in Figure \ref{fig:phasediagram}, with the only significant differences appearing for forward scattering interaction, with an $s$-wave superconducting phase rather than QAH at $|\lambda| > 0.2$ at $\mu = 0$, and a superconducting phase with degenerate $d_{xy}$ and $d_{x^2 - y^2}$ order parameters (leading to a chiral, ``$d+id$'' superconducting state) appearing at $\mu = 10^{-3} \frac{\Lambda^2}{2m}$ and $\lambda < -0.6$.

\begin{figure}
\includegraphics[width=0.45\textwidth]{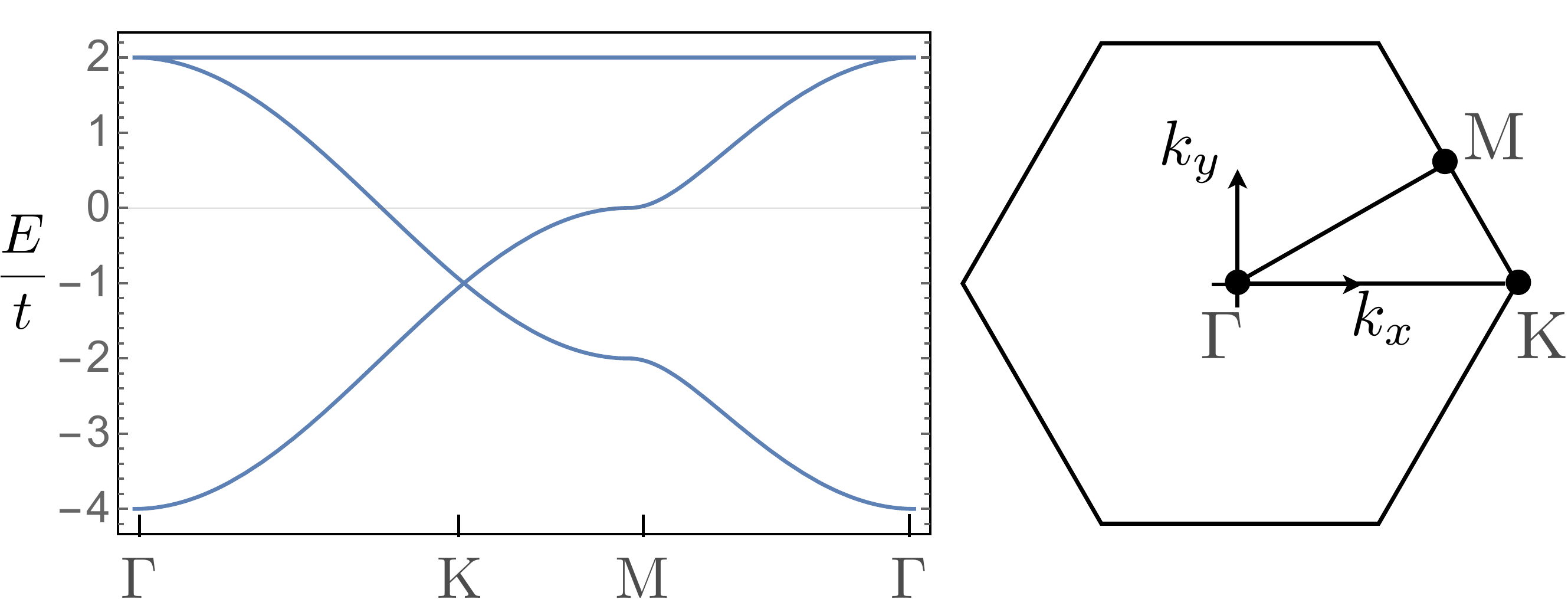}
\caption{Band structure for fermions on the kagome lattice with nearest-neighbor hopping $t$. A quadratic band touching point occurs between the upper and middle bands at the $\Gamma$ point ($\bk = 0$).}
\label{fig:1111b}
\end{figure}

Finally, let us consider the case of the kagome lattice as an example of a $C_{6v}$-symmetric system, in which case $\eta$ is fixed to $\frac{\pi}{4}$. A tight-binding calculation with nearest-neighbor hopping leads to three energy bands, with a completely flat upper band touching a parabolically dispersing middle band at $\bk = 0$, as shown in Figure \ref{fig:1111b}.
Thus, the system near $\frac{2}{3}$ filling corresponds to a QBC in the extreme particle-hole asymmetric case where $\lambda = -1/\sqrt{2}$. The two-band low energy effective theory introduced in Section \ref{sec:model} can be obtained by projecting out the completely filled lowermost band, after which the two components of the spinor $\psi_\sigma$ correspond to different linear combinations of the three fermion operators defined on the three sublattices. Starting with a microscopic Hubbard interaction in the lattice model and performing the projection onto the low-energy effective theory leads to the Hubbard interaction shown in Table \ref{table:interactions}, which is distinct from the form of the Hubbard interaction on the checkerboard lattice. Thus, although the two systems are described by the same effective field theory (at least in the case where $t_x = t_z$), knowledge of the original lattice model is retained through the form of the interactions. In fact, this can even affect which phase is realized. For the $C_{6v}$-symmetric system at $\mu=0$, Hubbard interaction leads to a QSH phase for $|\lambda| < 0.98 / \sqrt{2}$ and QAH for $0.98/\sqrt{2} < |\lambda| < 1/\sqrt{2}$, which is very different from the behavior shown in Figure \ref{fig:phasediagram}(a). In the case with one nearly flat band ($|\lambda| \lesssim 1/\sqrt{2}$), the couplings $\tilde g_i (\ell)$ turn out to all remain positive until very large values of $\ell$, with $C_1 \equiv \ell_1 g \gtrsim 10$. According to the condition \eqref{eq:1104a}, then, superconductivity is only to be expected in a vanishingly small range of $\mu$ for any bare interaction strength $g \lesssim 1$. Rather than leading to superconductivity, the coupling flows terminate when the UV cutoff reaches the Fermi energy, and the resulting phase in this case is a Fermi liquid.

\section{Conclusion}
While the appearance of superconductivity adjacent to various forms of particle-hole order has been observed in many families of strongly correlated materials, the precise mechanism for such behavior is still a matter of controversy. The fact that such phases tend to appear together appears to be at odds with the naive expectations of mean-field theory, from which one would expect such phases to compete, so that what is good for superconductivity is bad for particle-hole order and vice versa. The results of the preceding section suggest that these phases may be more usefully thought of as ``intertwined'' rather than competing,\cite{intertwined} as it is the enhanced logarithmic RG flows due to the fluctuations in particle-hole channels that ultimately lead to superconductivity. Of particular interest in the single-valley QBC model studied here is the unusual appearance of $s$-wave superconductivity driven by repulsive interactions, as well as the appearance of superconductivity adjacent to topological phases of matter, as shown in Figures \ref{fig:gamma} and \ref{fig:phasediagram}. Our work raises the possibility that there may exist a quantum critical point separating these phases, although one cannot rule out the possibility of, \textit{e.g.}, a first-order phase transition using the present approach. More broadly, the QBC model studied here thus provides a relatively simple and well-controlled arena in which to better understand the behavior of intertwined orders in 2D. 


This work was supported by the NSF CAREER award
under Grant No. DMR-0955561 (OV), NSF Cooperative
Agreement No. DMR-0654118, NSF  Graduate Research Fellowship under Grant No. DGE 1144085 (KP), and the State of Florida
(OV,JM).

%

\begin{appendix}
\section{Flow equation coefficients}
Let the bare part of the action be given by
\ba
\label{eq:1012c}
S_0 = T & \sum_n \int_0^{2\pi}\frac{d\theta_\bk}{2\pi} \int_0^{\Lambda(\theta_\bk)} \frac{dk}{2\pi} k \\
& \times \sum_\alpha \psi^\dagger_{n\bk\alpha} [ -(i\omega_n + \mu) 1_2 + \mathcal{H}_0 (\bk) ]
	\psi_{n\bk\alpha}, 
\ea
where we have taken the the UV cutoff to be dependent on the angle in $\bk$-space. We can write this as $\Lambda(\theta_\bk) = \Lambda f(\theta_\bk)$, and will choose $f(\theta_\bk)$ below.
We shall assume in what follows that $\mu \geq 0$, so that the Fermi level is at positive energy. In performing an RG step, we decrease the cutoff magnitude as $\Lambda \to \Lambda e^{-\ell}$, without scaling the angle-dependent part $f(\theta_\bk)$. Similarly, we scale the magnitude of all momenta as $k \to k e^\ell$, while leaving $\theta_\bk$ untouched. With the scaling for $\mu$, $\omega_n$, and $\psi_{n\bk\sigma}$ remaining the same as before, one sees that the bare action \eqref{eq:1012c} remains invariant.

The Green function obtained from \eqref{eq:1012c} is
\begin{widetext}
\begin{align}
\label{eq:1012e}
\hat{G}_0 (i\omega, \bk) = \frac{-(i\omega + \mu - \frac{\lambda}{\sqrt{2}m}k^2)1_2 
	- \frac{k^2}{\sqrt{2}m} ( \sin \eta \sin 2\theta_\bk \sigma_1 + \cos \eta \cos 2\theta_\bk \sigma_3)}
	{(i\omega + \mu)^2 
	- \left( \frac{k^2}{2m} \right)^2 
	\left(\sqrt{2}\lambda + \sqrt{1+\cos(2\eta) \cos(4\theta_\bk)} \right)^2}.
\end{align}
The outer product that is required in evaluating the one-loop diagrams is given by (taking the limit $T\to 0$)
\ba
\label{eq:1012f}
\int_>& \hat{G}_0 (+) \otimes \hat{G}_0 (\pm) \equiv 
	\int \frac{d\omega}{2\pi} \int_0^{2\pi} \frac{d\theta_\bk}{2\pi} 
	\int_{(1-d\ell)\Lambda f(\theta_\bk)}^{\Lambda f(\theta_\bk)} \frac{dk}{2\pi} k
	\hat{G}_0 (i\omega, \bk) \otimes \hat{G}_0 (\pm i\omega, \pm \bk) \\
&= \frac{\Lambda^2}{2\pi} d\ell \int \frac{d\omega}{2\pi} \int_0^{2\pi} 
	\frac{d\theta_\bk}{2\pi} f^2(\theta_\bk) \\
	&\times\frac{\left( i\omega+\mu - \frac{\lambda \Lambda^2}{\sqrt{2}m} \right)
	\left( \pm i\omega+\mu - \frac{\lambda \Lambda^2}{\sqrt{2}m} \right) 1_2 \otimes 1_2  + 2\left( \frac{\Lambda^2}{2m} \right)^2 
	\cos^2\eta \cos^2 2\theta_\bk \sigma_3 \otimes \sigma_3
	+ 2\left( \frac{\Lambda^2}{2m} \right)^2 
	\sin^2\eta \sin^2 2\theta_\bk \sigma_1 \otimes \sigma_1}
	{\left[ (i\omega+\mu-\frac{\Lambda^2}{2m}\sqrt{2}\lambda)^2 
	- \left( \frac{\Lambda^2}{2m}\right)^2 \left(1+\cos(2\eta) \cos(4\theta_\bk) \right) \right] 
	\left[ ( \pm i\omega+\mu-\frac{\Lambda^2}{2m}\sqrt{2}\lambda)^2 
	- \left( \frac{\Lambda^2}{2m}\right)^2 \left(1+\cos(2\eta) \cos(4\theta_\bk) \right) \right] } .
\ea
Below we consider \eqref{eq:1012f} separately for the particle-hole and particle-particle cases, which correspond to the upper and lower signs in the above equation, respectively.

For the particle-hole case, we find that the dependence on $\mu$, $\lambda$, and $f(\theta_\bk)$ disappears upon integrating over frequencies:
\ba
\label{eq:1012g}
\int_> \hat{G}_0 (+) \otimes \hat{G}_0 (+) = \frac{m}{4\pi} d\ell \int \frac{d\theta_\bk}{2\pi} 
	\bigg\{ & -\frac{1}{\sqrt{1+\cos(2\eta) \cos(4\theta_\bk)}} 1_2 \otimes 1_2 \\
& \quad + 2\frac{ \cos^2(\eta) \cos^2 (2\theta_\bk) \sigma_3 \otimes \sigma_3 + 
	\sin^2(\eta) \sin^2 (2\theta_\bk) \sigma_1 \otimes \sigma_1 }
	{[1+\cos(2\eta) \cos(4\theta_\bk)]^{3/2}} .
	\bigg\}
\ea
The angular integrals in this expression can be performed using special functions, leading to the following result:
\begin{align}
\label{eq:1012h}
\int_> \hat{G}_0 (+) \otimes \hat{G}_0 (+) = \frac{m}{4\pi} d\ell \left[ 
	-A^{(ph)} 1_2 \otimes 1_2 
	+ \frac{1}{2}B^{(ph)} \sigma_3 \otimes \sigma_3
	+ \frac{1}{2}C^{(ph)} \sigma_1 \otimes \sigma_1 \right],
\end{align}
where 
\begin{align}
\label{eq:1012i}
A^{(ph)}(\tilde\mu, \lambda, \eta) &= \frac{2}{\pi\sqrt{1+c_\eta}} K\left( \frac{2c_\eta}{1+c_\eta} \right) \\
B^{(ph)}(\tilde\mu, \lambda, \eta) &= (1+c_\eta) 
	\left[ _2 F_1 \left( \frac{3}{4}, \frac{5}{4}; 1; c_\eta^2 \right)
	- \frac{3}{4} c_{\eta _2} F_1 \left( \frac{5}{4}, \frac{7}{4}; 2; c_\eta^2 \right) \right] \\
C^{(ph)}(\tilde\mu, \lambda, \eta) &= (1-c_\eta) 
	\left[ _2 F_1 \left( \frac{3}{4}, \frac{5}{4}; 1; c_\eta^2 \right)
	+ \frac{3}{4} c_{\eta _2}  F_1 \left( \frac{5}{4}, \frac{7}{4}; 2; c_\eta^2 \right) \right].
\end{align}

We now turn to the particle-particle version of \eqref{eq:1012f}, defining the following coefficients: 
\begin{align}
\label{eq:1012m}
\int_> \hat{G}_0 (+) \otimes \hat{G}_0 (-) = \frac{m}{4\pi} d\ell \left[ 
	A^{(pp)}(\tilde\mu, \lambda, \eta) 1_2 \otimes 1_2 
	+ \frac{1}{2}B^{(pp)}(\tilde\mu, \lambda, \eta) \sigma_3 \otimes \sigma_3
	+ \frac{1}{2}C^{(pp)}(\tilde\mu, \lambda, \eta) \sigma_1 \otimes \sigma_1 \right].
\end{align}
In the most general case, the angular integrals in \eqref{eq:1012i} must be evaluated numerically, which greatly increases the computational cost of solving the flow equations. In light of this, we consider two special cases. In the first case, we do not assume particle-hole symmetry but do assume rotational invariance, so that $\eta = \frac{\pi}{4}$, but $\lambda$ is arbitrary. In this case we simply choose $f(\theta_\bk) = 1$, so that the UV cutoff has no angular dependence. Performing the integrals in \eqref{eq:1012i}, one obtains
\begin{align}
\label{eq:1012p}
A^{(pp)}(\tilde\mu, \lambda, \frac{\pi}{4}) = B^{(pp)}(\tilde\mu, \lambda, \frac{\pi}{4}) =  
	C^{(pp)}(\tilde\mu, \lambda, \frac{\pi}{4}) = \frac{1}{(1+\sqrt{2}\lambda)^2 - \tilde\mu^2 } .
\end{align}

In the second case we allow for angular anisotropy, but require particle-hole symmetry ($\lambda$=0). In order that the UV cutoff describes a contour of constant energy, we choose in this case \hbox{$f(\theta_\bk) = 1/[1+\cos(2\eta)\cos(4\theta_\bk)]^{1/4}$}, so that $\varepsilon_+ (\bk) |_{\Lambda f(\theta_\bk)} = \Lambda^2/2m$, which is independent of $\theta_\bk$. Then performing the angular integrals from \eqref{eq:1012i}, one obtains the following result:
\begin{align}
\label{eq:1012n}
A^{(pp)}(\tilde\mu, 0, \eta) &= \frac{2}{\pi\sqrt{1+c_\eta}} \frac{1}{1-\tilde\mu^2}
	K\left( \frac{2c_\eta}{1+c_\eta} \right) \\
B^{(pp)}(\tilde\mu, 0, \eta) &= \frac{1+c_\eta}{1-\tilde\mu^2} 
	\left[ _2 F_1 \left( \frac{3}{4}, \frac{5}{4}; 1; c_\eta^2 \right)
	- \frac{3}{4} c_{\eta _2} F_1 \left( \frac{5}{4}, \frac{7}{4}; 2; c_\eta^2 \right) \right] \\
C^{(pp)}(\tilde\mu, 0, \eta) &= \frac{1-c_\eta}{1-\tilde\mu^2} 
	\left[ _2 F_1 \left( \frac{3}{4}, \frac{5}{4}; 1; c_\eta^2 \right)
	+ \frac{3}{4} c_{\eta _2} F_1 \left( \frac{5}{4}, \frac{7}{4}; 2; c_\eta^2 \right) \right].
\end{align}
\end{widetext}
In the case where we assume rotational invariance and particle-hole symmetry such that $\eta=\frac{\pi}{4}$ and $\lambda=0$, we obtain from \eqref{eq:1012i} and either \eqref{eq:1012n} or \eqref{eq:1012p} the following coefficients:
\begin{align}
A^{ph}=B^{ph}=C^{ph} &=1, \\
A^{pp}=B^{pp}=C^{pp} &=\frac{1}{1-\tilde\mu_\ell^2},
\end{align}
which corresponds to the case analyzed in the main text.

\begin{figure}
\centering
\includegraphics[width=0.45\textwidth]{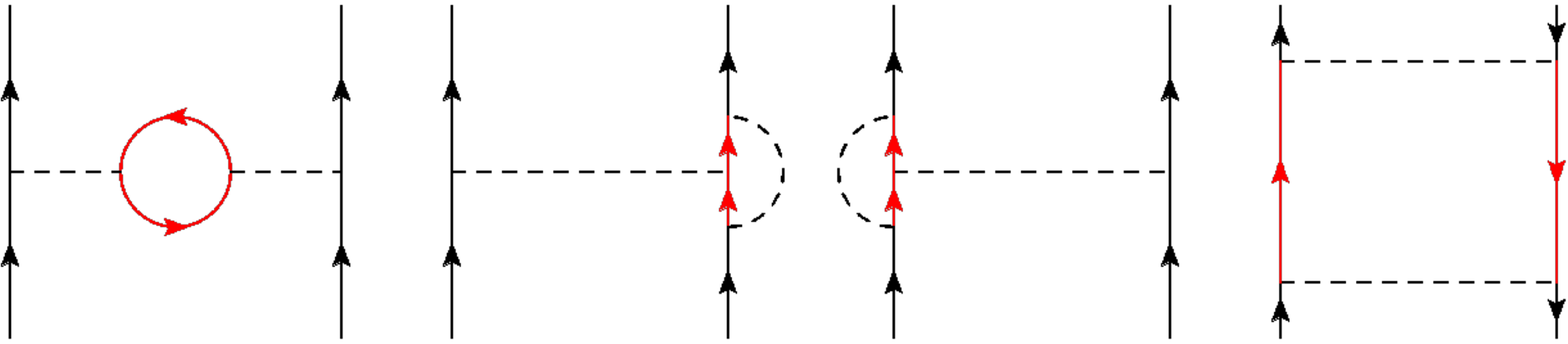}
\caption{Particle-hole diagrams contributing to the one-loop flow equations for the couplings $g_i(\ell)$.The solid lines denote fermion propagators, the dashed lines denote interactions, and the red lines denote fermion modes to be integrated over all frequency and momenta with $(1-d\ell)\Lambda<|\bk|<\Lambda$.}
\label{fig:4diagram}
\end{figure}

\begin{figure}
\centering
\includegraphics[width=0.1\textwidth]{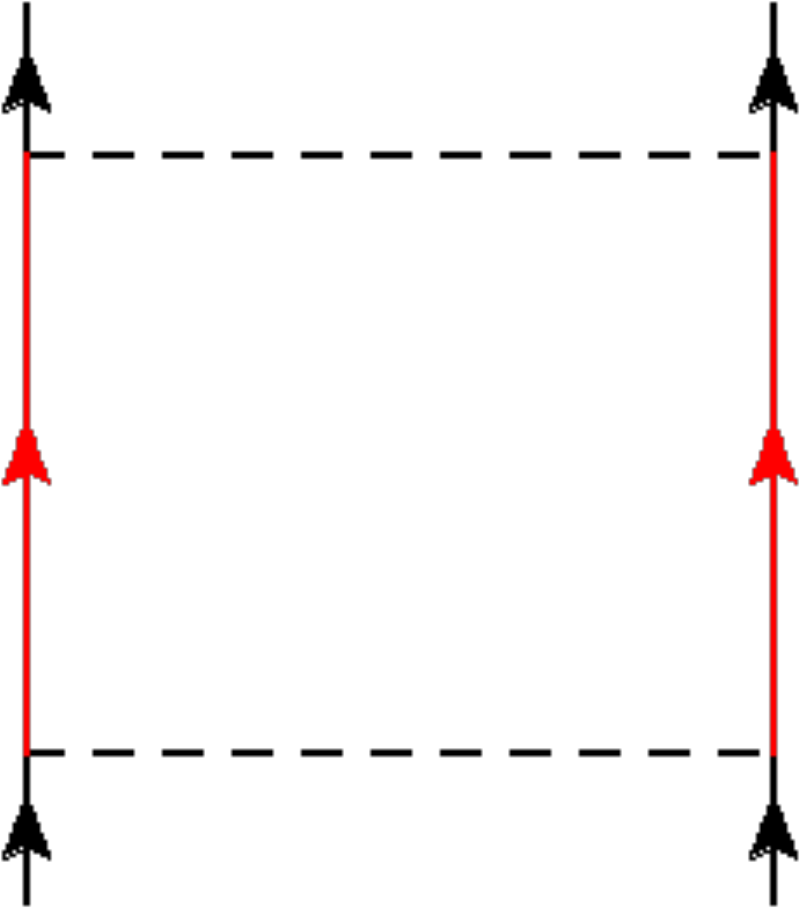}
\caption{Particle-particle diagram contributing to the one-loop flow equations for the couplings $g_i(\ell)$}\label{foo}
\label{fig:ppdiagram}
\end{figure}

With these results, we can proceed to calculate the coefficients in the flow equations. There are five marginally relevant one-loop diagrams which contribute to the coupling flows \eqref{eq:0128a}, shown in Figures \ref{fig:4diagram} and \ref{fig:ppdiagram}. Let $A_{ijk} = \sum_{d=1}^5 A^5_{ijk}$, where $d$ corresponds to one of these five diagrams. 
The first diagram only gives diagonal contributions:
\ba
A_{iii}^1 =  \big[ & -4 A^{ph}
	+ C^{ph}~\mathrm{Tr} ((\sigma_1 \sigma_j)^2) \\
	&+ B^{ph}~\mathrm{Tr} ((\sigma_3 \sigma_j)^2) \big] \frac{m}{4\pi}.
\ea
From the second and third diagrams combined:
\ba
\label{eq:0128d}
& A_{iij}^2 + A_{iij}^3 =  \frac{m}{4\pi} \bigg[ 
	A^{ph}~\mathrm{Tr}(\sigma_i \sigma_j\sigma_i \sigma_j) \\
& - \frac{1}{2}C^{ph}~\mathrm{Tr} (\sigma_i \sigma_j\sigma_1 \sigma_i \sigma_1 \sigma_j) 
	- \frac{1}{2}B^{ph}~\mathrm{Tr} 
	(\sigma_i \sigma_j\sigma_3 \sigma_i \sigma_3 \sigma_j) \bigg] .
\ea
From the fourth diagram:
\ba
A_{ijk}^4 = \frac{m}{32\pi} \bigg[ & 2A^{ph}
	~\mathrm{Tr}(\sigma_k \sigma_j \sigma_i)~\mathrm{Tr}(\sigma_j \sigma_k \sigma_i) \\
&- C^{ph}~\mathrm{Tr}(\sigma_k \sigma_1 \sigma_j \sigma_i)
	~\mathrm{Tr}(\sigma_j \sigma_1 \sigma_k \sigma_i) \\
&- B^{ph}~\mathrm{Tr}(\sigma_k \sigma_3 \sigma_j \sigma_i)
	~\mathrm{Tr}(\sigma_j \sigma_3 \sigma_k \sigma_i) \bigg]  .
\ea
And from the fifth diagram:
\begin{equation}
\begin{aligned}
A_{ijk}^5 = -\frac{m}{32\pi} \bigg[ & 2A^{pp}
	\left(\mathrm{Tr}(\sigma_k \sigma_j \sigma_i)\right)^2 \\
	&+ C^{pp} \left(\mathrm{Tr}(\sigma_k \sigma_1 \sigma_j \sigma_i)\right)^2 \\
	&+ B^{pp} \left(\mathrm{Tr}(\sigma_k \sigma_3 \sigma_j \sigma_i)
	\right)^2 \bigg] 
\end{aligned}
\end{equation}

\begin{figure}
\centering
\includegraphics[width=0.45\textwidth]{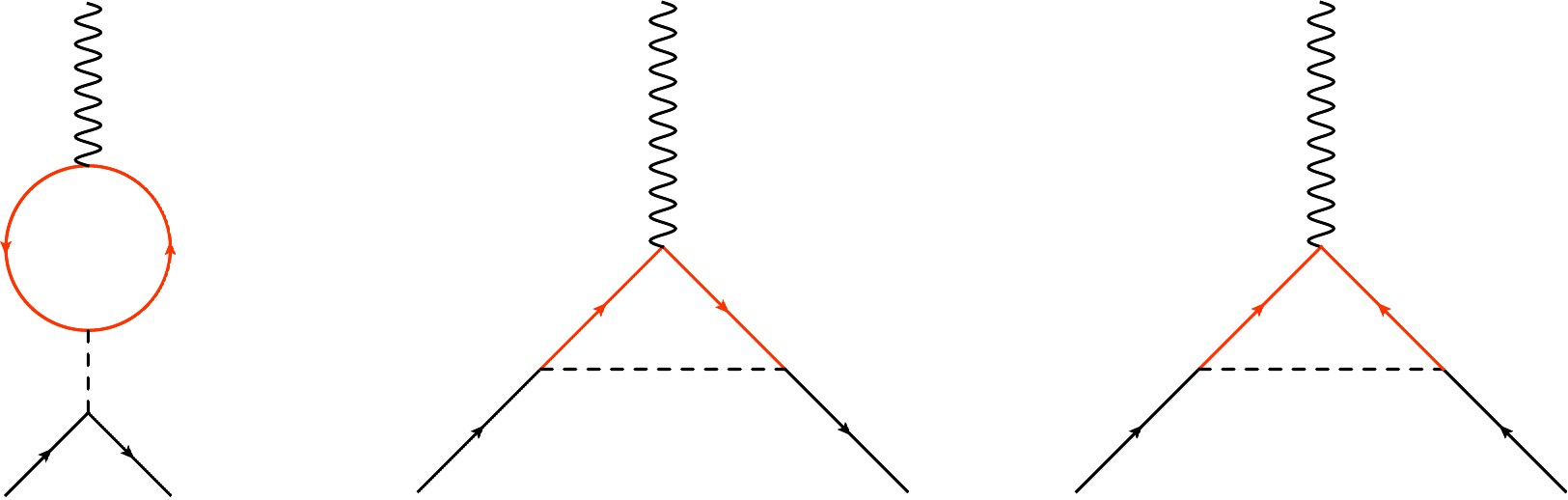}
\caption{One-loop diagrams representing contributions to the source terms $\Delta_i^{c,s,pp} (\ell)$.}\label{foo}
\label{fig:3diagram}
\end{figure}

The flows of the source terms introduced in \eqref{eq:0206a} are computed by evaluating the diagrams
shown in Figure \ref{fig:3diagram}.
Evaluating these diagrams gives the following expressions for the coefficients appearing in \eqref{eq:1103b}:
\begin{widetext}
\begin{equation}
\begin{aligned}
\label{eq:0220c}
B_{ij}^{(c,s)}(1) &= \frac{m}{4\pi} \{ - A^{ph} \mathrm{Tr} [(\sigma_j 1) M_i^{(c,s)}] 
	+ \frac{1}{2} B^{ph} \mathrm{Tr} [(\sigma_j 1) (\sigma_3 1) M_i^{(c,s)} (\sigma_3 1)]
	+ \frac{1}{2} C^{ph} \mathrm{Tr} [(\sigma_j 1) (\sigma_1 1) M_i^{(c,s)} (\sigma_1 1)] \}, \\
\end{aligned}
\end{equation}
\begin{equation}
\begin{aligned}
B_{ij}^{(c,s)}(2) &= -\frac{m}{16\pi} \{ -A^{ph} \mathrm{Tr} [((\sigma_j 1) M_i^{(c,s)})^2] 
	+ \frac{1}{2}B^{ph} \mathrm{Tr} [M_i^{(c,s)} (\sigma_j 1) (\sigma_3 1) M_i^{(c,s)} (\sigma_3 1) (\sigma_j 1)] \\
	&\quad\quad\quad\quad\quad\quad\quad\quad\quad\quad
	+ \frac{1}{2}C^{ph} \mathrm{Tr} [M_i^{(c,s)} (\sigma_j 1) (\sigma_1 1) M_i^{(c,s)} (\sigma_1 1) (\sigma_j 1)] \}, \\
\end{aligned}
\end{equation}
\begin{equation}
\begin{aligned}
B_{ij}^{(pp)} &= -\frac{m}{16\pi} \{ A^{pp} \mathrm{Tr} [(\sigma_j 1) M_i^{(pp)} (\sigma_j 1)^T  M_i^{(pp)}] 
	+ \frac{1}{2}B^{pp} \mathrm{Tr} 
	[ M_i^{(pp)} (\sigma_j 1) (\sigma_3 1)  M_i^{(pp)} (\sigma_3 1) (\sigma_j 1)^T] \\
	&\quad\quad\quad\quad\quad\quad\quad\quad\quad\quad
	+ \frac{1}{2}C^{pp} \mathrm{Tr} 
	[ M_i^{(pp)} (\sigma_j 1) (\sigma_1 1)  M_i^{(pp)} (\sigma_1 1) (\sigma_j 1)^T] \} .
\end{aligned}
\end{equation}
\end{widetext}
Adding the contributions from the first two diagrams together then gives $B^{(c,s)}_{ij} = B^{(c,s)}_{ij}(1) + B^{(c,s)}_{ij}(2)$. 
By differentiating the free energy with respect to the source terms, one obtains the following expression for the particle-hole susceptibilities in the charge and spin channels:
\ba
\label{eq:0220a}
&\chi_i^{(c,s)}(\ell) = \frac{m}{4\pi} \int_0^\ell d\ell' e^{2\Omega_i^{(c,s)}(\ell')}
	\bigg\{ 4 A^{ph}  \\
& \quad - \frac{1}{2} B^{ph} \mathrm{Tr} [((\sigma_3 1) M_i )^2]
	- \frac{1}{2} C^{ph} \mathrm{Tr} [((\sigma_1 1) M_i )^2] \bigg\},
\ea
where
\begin{align}
\label{eq:0220b}
\Omega_i^{(c,s)} (\ell) = \int_0^\ell d\ell' \sum_{j=0}^3 B_{ij}^{(c,s)} g_j(\ell').
\end{align}
The particle-particle susceptibilities are given by
\ba
\label{eq:0220a}
&\chi_i^{pp}(\ell) = \frac{m}{4\pi} \int_0^\ell d\ell' e^{2\Omega_i^{pp}(\ell')}
	\bigg\{ 4 A^{pp} \\
& \quad + \frac{1}{2} B^{pp} \mathrm{Tr} [((\sigma_3 1) M_i )^2]
	+ \frac{1}{2} C^{pp} \mathrm{Tr} [((\sigma_1 1) M_i )^2] \bigg\},
\ea
where
\begin{align}
\label{eq:0220b}
\Omega_i^{pp} (\ell) = \int_0^\ell d\ell' \sum_{j=0}^3 B_{ij}^{pp} g_j(\ell').
\end{align}

Having determined the complete flow equations for the couplings and susceptibilites, we can proceed to solve the coupled differential flow equations given initial conditions $g_i(0)$ and $\mu(0)$. We assume that $-\frac{1}{\sqrt{2}} < \lambda < \frac{1}{\sqrt{2}}$, so that the bands do not cross the Fermi level away from the point $\bk = 0$. The phase instability is taken to be the in the channel in which the susceptibility $\chi_i$ is the largest when the couplings reach values $|g_i(\ell)| \gtrsim 1$. The phase diagrams obtained in this way are shown in Figure \ref{fig:phasediagram}.
\\
\\
\end{appendix}


\begin{thebibliography}{99}
\bibitem{TrigWarpingFootnote}
In AB stacked bilayer honeycomb, further neighbor interlayer hopping leads to so-called ``trigonal warping,'' which splits the parabolically touching bands into four miniature Dirac cones in a way that preserves the threefold rotational symmetry. Symmetry-breaking phases can still occur if the electron interactions are sufficiently strong relative to the strength of the further neighbor hopping, as is thought to be the case in bilayer graphene.\cite{Throck,MurrayLong}
\bibitem{gammaFootnote}
Note that the quantity plotted in Figure 2 of Ref.~\onlinecite{Murray} corresponds to $\gamma_m^{c,s} +1$ in our present notation, where $\gamma_m^{c,s}$ is the susceptibility exponent defined in \eqref{eq:1103b}. Because the leading susceptibilities found in Ref.~\onlinecite{Murray} all have exponents $\gamma_m \geq 0$ (using our present definition of $\gamma_m$), the physical conclusions about the leading diverging susceptibilities remain unaffected.
\end{thebibliography}
\end{document}